\documentclass[aps,prl,twocolumn,superscriptaddress]{revtex4-1}
\usepackage{graphicx}
\usepackage{amsmath}

\begin{document}

\title{Validating Continuum Lowering Models via Multi-Wavelength Measurements of Integrated X-ray Emission}
\author{M.F. Kasim}
\affiliation{Department of Physics, Clarendon Laboratory,University of Oxford, Parks Road, Oxford OX1 3PU, UK}
\author{J.S. Wark}
\affiliation{Department of Physics, Clarendon Laboratory,University of Oxford, Parks Road, Oxford OX1 3PU, UK}

\author{S.M. Vinko}
\email{sam.vinko@physics.ox.ac.uk}
\affiliation{Department of Physics, Clarendon Laboratory,University of Oxford, Parks Road, Oxford OX1 3PU, UK}

\date{\today}

\begin{abstract}
X-ray emission spectroscopy is a well-established technique used to study continuum lowering in dense plasmas. It relies on accurate atomic physics models to robustly reproduce high-resolution emission spectra, and depends on our ability to identify spectroscopic signatures such as emission lines or ionization edges of individual charge states within the plasma. Here we describe a method that forgoes these requirements, enabling the validation of different continuum lowering models based solely on the total intensity of plasma emission in systems driven by narrow-bandwidth x-ray pulses across a range of wavelengths. The method is tested on published Al spectroscopy data and applied to the new case of solid-density partially-ionized Fe plasmas, where extracting ionization edges directly is precluded by the significant overlap of emission from a wide range of charge states.
\end{abstract}

\pacs{Valid PACS appear here}
\maketitle

\section{Introduction}

Continuum lowering (CL) is a fundamental physical process whereby the ionization potential of an ion embedded within a dense plasma is lowered due to the interaction of the ion and its electrons with the neighbouring plasma particles. As a process it is of paramount importance in the study and modelling of matter at high energy-densities (HED) common in systems of astrophysical and inertial confinement fusion relevance, because it alters all binding energies within a plasma, and modifies the electron equation of state, ionization balance, collisional dynamics and material transport properties.
Due to its broad importance, the effects of CL in dense plasmas have been investigated employing an extensive range of theoretical approaches, including average atom~\cite{PhysRevE.87.013104}, Debye screening, and ion-sphere models~\cite{PhysRevA.93.052513,PhysRevA.87.062502};
fundamental plasma calculations~\cite{Belkhiri2013,Crowley2014};
classical molecular dynamics~\cite{Calisti:2015} and Monte Carlo simulations~\cite{Stranski:2016};
quantum statistical theory~\cite{PhysRevE.96.013202};
and {\it ab initio} methods based on density functional theory~\cite{Vinko2014,PhysRevB.93.115114,PhysRevLett.119.065001}.

Progress has been slower on the experimental front due to difficulties in creating well-defined dense plasma systems, and in devising accurate diagnostics to reliably measure continuum lowering. The most successful diagnostics to date have been x-ray inelastic scattering and x-ray emission spectroscopy. X-ray scattering experiments have recently investigated compressed CH plasmas on large-scale laser facilities Omega~\cite{Fletcher2014} and the NIF~\cite{PhysRevE.94.011202}, while spectroscopy has been deployed in campaigns investigating Mg, Al and Si using x-ray isochoric heating at the Linac Coherent Light Source (LCLS) free-electron laser~\cite{Ciricosta2012,Ciricosta:2016aa,PhysRevLett.119.085001}, and in fundamental plasma experiments on Al on the Orion laser at AWE~\cite{Hoarty2013}.

Spectroscopic methods are based on the strong dependence that the atomic physics of ions in a plasma exhibits to CL effects. This dependence makes it possible to identify specific features in x-ray emission and absorption spectroscopy, such as bound-bound transition lines~\cite{Preston2013} or the presence of absorption or emission edges, that can be linked, directly or indirectly, to the ionization threshold energy, and hence to the level of CL.
Inevitably, the method also relies on the availability of predictive atomic kinetics simulations with accurate atomic physics models to robustly reproduce high-resolution emission spectra.

The experimental work to date has been centred around materials with low atomic number Z, and there is no reliable data for plasmas formed of highly ionized, dense, heavier elements. However, the effect of CL increases both with increasing density and ionization (e.g., see the analytical models in ref.~\cite{Stewart:1966ay}), so that the largest effects will present in highly ionized, high-Z materials. These systems pose several important difficulties to spectroscopic investigations in its current form. Firstly, as the atomic number increases it becomes ever more difficult to identify spectroscopic features attributable to a specific charge state, as the growing number of weakly bound electrons means that many different charge states emit at similar wavelengths. This overlap can make it difficult to extract ionization edges or identify desired bound-bound transitions. So while the total spectroscopic signature may well continue to encode information on the physics of CL, extracting it requires complex spectral modelling even in the absence of other experimental uncertainties such as gradients in the plasma conditions. The second important difficulty is that accurate spectral modelling of partially-ionized systems is increasing difficult in its own right as Z increases, even without the added complication of plasma effects, given the need to account for the rapidly increasing number of transitions. Finally, any plasma modelling needs to include some CL model for comparison with experiment, and it is not immediately obvious that there is a single combination of plasma conditions, atomic physics model and CL model that will provide a unique agreement with the experimental result.

Here we show that the total integrated x-ray emission intensity of a system, driven by a bright and narrow-bandwidth x-ray pulse to HED conditions, depends predictably on the degree of CL in the plasma. By varying the wavelength of the driving x-ray pulse, at the appropriate intensity, it is possible to identify large absolute and relative differences in the emitted intensity, and to discriminate between CL models without the need to identify spectral features, or indeed measure the spectrally resolved emission at all. The method makes use of the CL dependence of the number of states present in the system capable of absorbing and emitting x-rays, and can be modelled well by relatively simple atomic kinetics simulations. We propose that by measuring the integrated emission intensity as a function of excitation photon energy it is thus possible to validate specific CL models in a way that is both experimentally practical and computationally efficient.

\section{Approach and validation in Al}

\begin{figure}
\includegraphics[width=\linewidth]{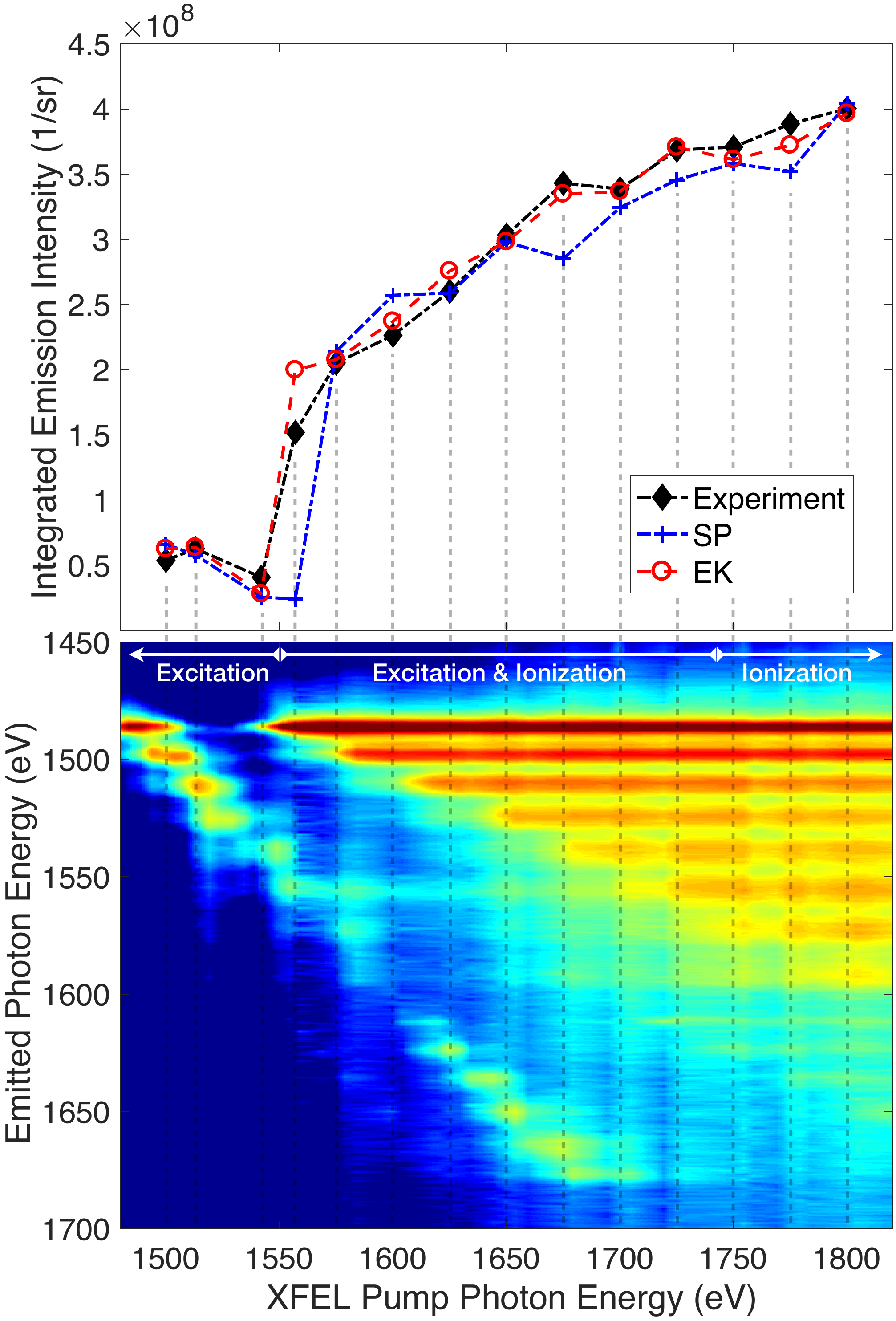}
\caption{\label{fig1} Emitted x-ray intensity from an Al system driven by an FEL pulse centred at the specified pump photon energies. The experimental values from refs.~\cite{Vinko2012,Cho2012} are compared with SCFLY atomic kinetics modelling~\cite{Ciricosta:2016POP} using the Stewart-Pyatt (SP) and Ecker-Kr\"oll (EK) continuum lowering models. Three regions based on whether the system is primarily driven by excitation, ionization, or both, are identified.}
\end{figure}

Recently reported isochoric heating experiments performed at the LCLS free-electron laser (FEL)~\cite{Vinko2012,Cho2012,Vinko:jpp2015} provide a rich spectroscopy dataset for x-ray-driven aluminium plasmas that we shall employ for the validation of our proposed approach. 
In these experiments, a monochromatic x-ray pulse of well-defined photon energy is focused to spots of $\sim$10~$\mu$m$^2$ on thin foils of various materials. Typical intensities achieved are on the order of 10$^{17}$~Wcm$^{-2}$, sufficient to heat the irradiated regions to temperatures exceeding 100~eV on femtosecond timescales, and to drive resonant and non-linear atomic processes. The intense x-ray pulse can drive x-ray photoionization of inner-shell electrons, provided the photon energy is higher than the shell's ionization edge, and bound-bound transitions leading to excited atomic configurations, if the resonance energies are within the bandwidth of the x-ray pulse. Recombination into the core holes created by this interaction produces strong x-ray emission that has been spectrally resolved to identify ionization edges and deduce levels of CL~\cite{Ciricosta2012,Ciricosta:2016aa}, observe non-linear processes~\cite{Cho2012}, measure x-ray opacities~\cite{PhysRevLett.119.085001} and extract electron collisional ionization rates~\cite{Vinko2015,quincy:2017}.

\begin{figure*}[t]
\includegraphics[width=\linewidth]{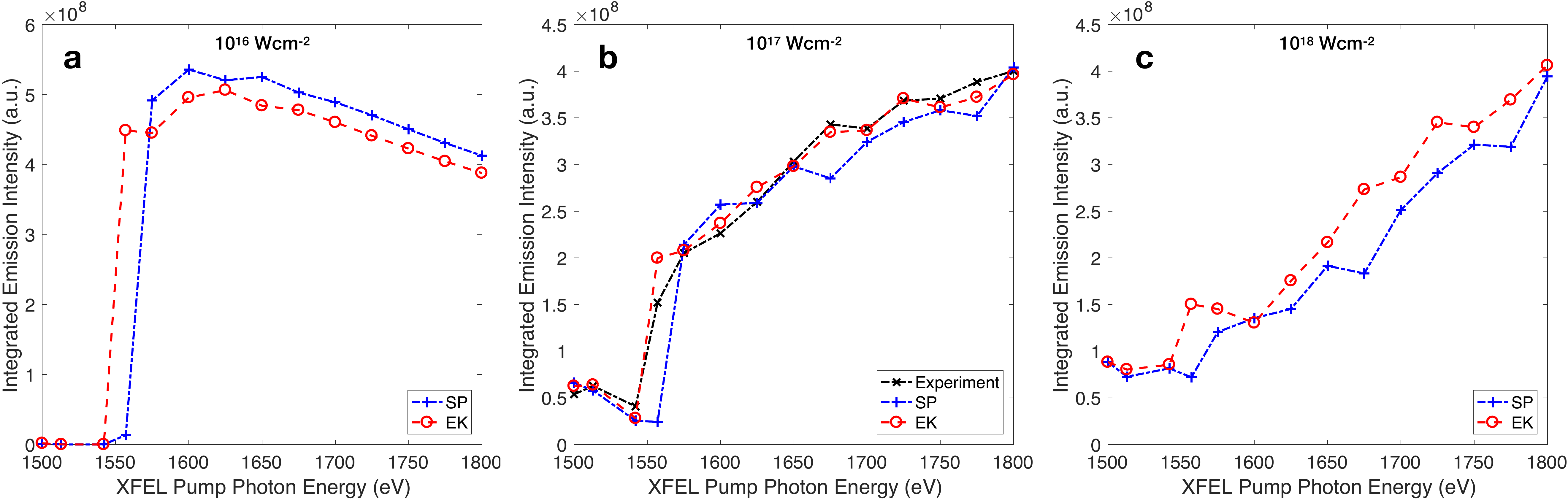}
\caption{\label{fig2} Total emitted x-ray intensity from Al driven by an FEL pulse centred at the specified pump photon energies. Panel~b contains the same data as Fig.\ref{fig1}. The intensities correspond to peak intensities of a supergaussian intensity distribution, describing a real experimental measurement.}
\end{figure*}

For the purposes of the work here we will only be interested in the total x-ray emission intensity, integrated in wavelength. Specifically for the case of Al present in the literature, we integrate the spectrally-resolved experimental data over the range of K$_{\alpha}$ emission wavelengths for FEL photon energies between 1500~eV and 1800~eV. We plot these intensities in Fig.~\ref{fig1}, alongside atomic kinetics modelling predictions using the SCFLY code~\cite{SCFLYcode}. The modelling follows the methods discussed in ref.~\cite{Ciricosta:2016POP}, and includes either the Ecker-Kr\"oll CL model (EK, with C=1)~\cite{EckerG.Kroll1963,Ciricosta2012,Ciricosta:2016aa,Preston2013}, or the Stewart-Pyatt CL model (SP)~\cite{Stewart:1966ay}. The collisional cross sections given by van den Berg et al.~\cite{quincy:2017} are used throughout. The simulations have been scaled to match the experimental data at the highest pump energy. The experimental integrated emission is plotted in units of emitted X-ray photons per solid angle. Absolute intensities were calculated taking into account both experimental geometry and spectrometer throughput, and are accurate to $\sim$10\%.

We identify three specific regions of x-ray pump photon energies, based on the relative importance of photoionization and photoexcitation in driving the x-ray-matter interaction.
The first region is that where photoexcitation dominates, i.e., where the FEL pulse pumps the system at or near a resonant bound-bound transition. Here this corresponds to a $1s-2p$ transition in an Al ion, and the photon energy of the FEL is just below that required for direct K-shell ($1s$) photoionization (1560~eV). Emission in this region is generally expected to be weak because the absorption cross section to the pump x-rays is low unless the system is at resonance, and getting to the resonance condition requires a considerable energy density to have been deposited in the sample via inefficient off-resonance processes.

The second region is the mixed region present when the FEL pump is tuned above the K-edge of the neutral Al atom (cold K-edge), but remains below the highest available bound-bound transition (in the absence of $n=3$ states, the Ly$_{\alpha}$ transition at $\sim$1.73~keV). Here both K-shell photoionization and photoexcitation can drive absorption and emission. Higher pump energies allow more ions to interact with the x-ray pulse, so the emission generally increases with increasing photon energy. This trend is promptly confirmed by the experimental data and both simulations.

\begin{figure*}[t]
\includegraphics[width=\linewidth]{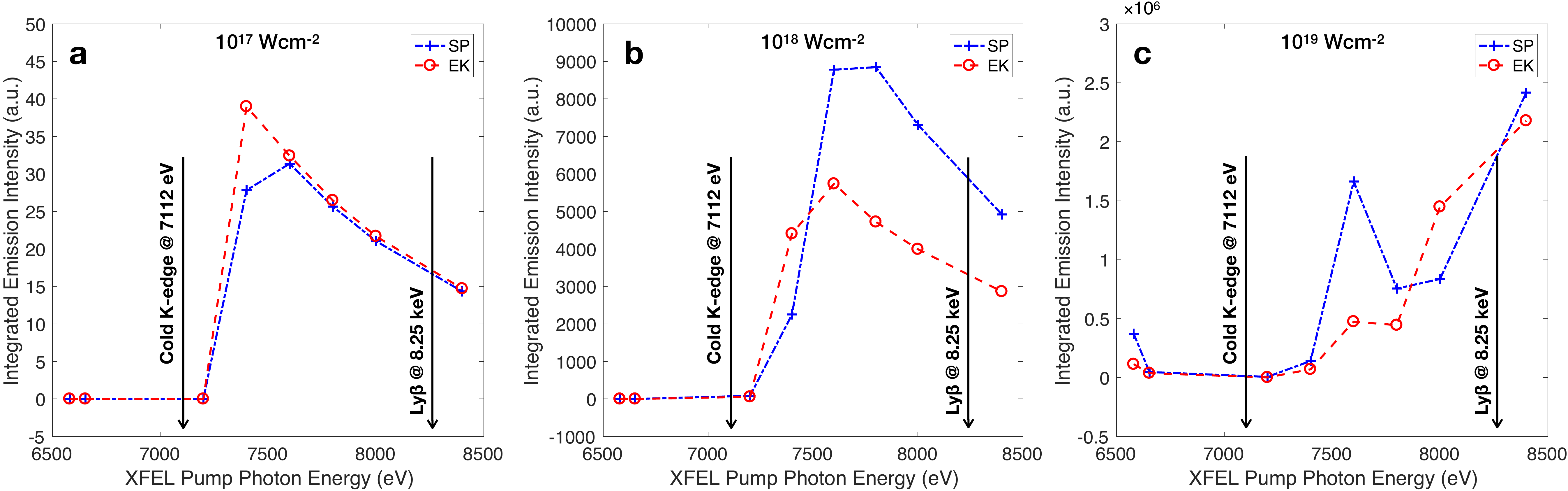}
\caption{\label{fig3} Simulated total emitted x-ray intensity from Fe driven by an FEL pulse centred at the specified pump photon energies. The Stewart-Pyatt (SP) and Ecker-Kr\"oll (EK) continuum lowering models are compared. The intensities correspond to peak intensities of a supergaussian intensity distribution, simulating a real experimental measurement. The photon energy corresponding to peak emission is correlated with the highest charge state that can be thermally generated, and is seen to increase from around 7.4 keV at 10$^{17}$~Wcm$^{-2}$, to over 8.4~keV at 10$^{19}$~Wcm$^{-2}$.}
\end{figure*}

The third region is where the XFEL pulse is tuned above all bound-bound transition energies in the system. Here, most states present in the plasma can be photoionized, so the number of states participating in the interaction grows more slowly and eventually becomes constant as the pump photon energy increases. However, since above-edge absorption cross sections decrease with increasing photon energy, the total emission is expected to tail off and eventually start decreasing. While we do have experimental data for X-ray pumping above the Ly$_{\alpha}$ energy, data for X-ray pumping above the H-like K-edge is not presently available, so that only a tailing off of the intensity is observed in Fig.~\ref{fig1}, but not the full inversion.

The two CL models, SP and EK, are indistinguishable for the lowest energy points in region I where the system is resonantly pumped. This is unsurprising as the difference in predicted IPD is small for the first few charge states. For higher resonances the effects start to overlap with direct K-shell photoionzation and become harder to isolate, but there remains a marked difference in CL models around the cold-K-edge. This is in part due to the small difference in predicted K-edge energy combined with FEL jitter and bandwidth, and in part to the larger overall resonance emission in the EK model where the resonance charge state is more easily reached.

At pump energies above the cold ionization edge the emission intensity is seen to increase monotonically with pump photon energy. This is reproduced in the simulations by both CL models, albeit with different slopes. In particular, the slope of increase within the SP model is smaller than that calculated according to the EK model, indicative of the lower predicted values of CL. 
Further, because the SP model predicts less CL, the highest ionization threshold will be larger than in calculations with the EK model, requiring higher photon energies to interact with the full range of plasma charge states present in the system. The inversion from an increasing to a decreasing intensity as a function of photon energy is therefore another useful integrated measure to evaluate the degree CL. Unfortunately, the present data does not extend sufficiently high in photon energy to make this comparison possible here.

Overall, the slope of the data favours the EK model, an observation that, notably, does not depend on the identification of any specific ionization threshold. We note that this same region is where features in the spectrum are used by Ciricosta et al. to extract ionization edges and IPDs~\cite{Ciricosta2012,Ciricosta:2016aa}, an approach contested by Iglesias on the basis that more detailed atomic modelling could perhaps lead to different answers~\cite{Iglesias2014}.

Across all the points shown in Fig.~\ref{fig1}, the average difference between the measured intensities and those predicted via the EK model is around 7\%, and is just over twice that for SP. With a few exceptions, the intensity differences for a single point are therefore small and would be challenging to measure with sufficient accuracy in experiments for model validation. We note, however, that the emission intensity has a notable intensity dependence. We show in Fig.~\ref{fig2} the integrated emission for Al irradiated at intensities between 10$^{16}$~Wcm$^{-2}$ and 10$^{18}$~Wcm$^{-2}$. Figure~\ref{fig2}b contains the same data as Fig.~\ref{fig1}. We note that by increasing the intensity up to 10$^{18}$~Wcm$^{-2}$, the differences in the predicted emission from the SP and EK models increases significantly -- by up to a factor two -- well above the levels required for experimental discrimination. Selecting the appropriate intensity is thus seen to play an important role in the application of the total emission measurement to CL model validation.

\section{Continuum Lowering in Fe}

There is much interest in exploring the physics of continuum lowering in higher Z materials and at higher ionizations. However, the methods employed for CL investigations in lower-Z material such as Mg, Al and Si on the LCLS FEL~\cite{Ciricosta:2016aa} are unlikely to be suitable for heavier ions as the significant overlap of emission from a plethora of charge states makes it challenging to identify emission features attributable to specific charge states. In particular, it is uncertain whether clear ionization thresholds can be observed experimentally. Of course, changes in the spectral emission as a function of the photon energy of the x-ray pump can still be compared with synthetic spectra and used to extract physical parameters, but this approach will depend strongly on the detailed accuracy of the synthetic spectral model. Moreover, for higher Z elements such as Fe it is unlikely that a superconfigurational atomic kinetics model will be sufficient as a basis on which to conduct spectral modelling, and the number of configurations (or indeed, states) required for the modelling may well make such calculations, even if feasible, very computationally expensive. It is thus interesting to explore whether the total integrated emission intensity can be used as a proxy for CL model validation. In particular, if significant relative differences in the total emission can be observed between systems pumped away from absorption edges, the computational modelling requirements could be relaxed making the conclusions more robust to potential inaccuracies and simplifications in atomic models.

We show a set of simulated Fe emission intensities in Fig.~\ref{fig3}. The modelling was done using the non-LTE SCFLY superconfigurational atomic kinetics code. The pump x-ray pulse is modelled by a Gaussian with a FWHM of 60~fs, and the calculation is run for 300~fs in 1~fs time steps. The plotted results are not single-intensity simulations, but assume a typical intensity distribution on target using the realistic capabilities of x-ray focusing at the LCLS facility, as measured previously via the f-scan technique~\cite{Chalupsky:2010bv,Chalupsky:2013}. The intensities quoted in Fig.~\ref{fig3} represent peak intensities, with the rest of the approximately supergaussian intensity distribution sampled using 50 calculations covering four orders of magnitude in intensity. The modelling thus accounts for both temporal and spatial variations of the FEL x-ray pulse, and should be representative of the actual experiment, much in line with the methods described in ref.~\cite{Ciricosta:2016POP}. The simulations are identical in all but the CL model.

We may again distinguish three x-ray photopumping regions, as was done previously for Al. The first region is again that dominated by photoexcitations, where $1s-2p$ and now also $1s-3p$ transitions are pumped on or near resonance, in the presence of higher-lying spectator electrons. This region is clearly dominant at photon energies below the cold Fe K-edge, but can extend higher for increasingly hot and ionized systems as the $1s$ ionization thresholds increase with increasing ionization. At photon energies above the cold Fe K-edge, K-shell x-ray photoionization becomes possible, and competes with photoexcitation processes amongst bound states in more highly ionized ions. This is the mixed, second region for Fe.
Finally, for photon energies above the He$_{\beta}$ transition ($1s-3p$), photoionization in highly ionized ions becomes the main x-ray-matter interaction process. We identify this as the photoionization region, tacitly ignoring for now the small contributions from double-core-hole ionization in highly-ionized Fe.

As alluded to in the Al simulations, the behaviour of the system is seen to strongly depend on the intensity of the x-ray pump. This is because significant energy densities are needed in order to highly ionize solid-density iron. In particular, we observe that it is far more challenging to produce bright $2p-1s$ resonantly pumped emission in Fe than in Al, because since the resonant state needs to be created collisionally, high temperatures in excess of some 500~eV are required. This is several times the temperature required to achieve a similar effect in Al.

From Fig.~\ref{fig3} it is clear that there is very little difference in the integrated emission from Fe at peak intensities around $10^{17}$~Wcm$^{-2}$. This is the intensity regime typically achieved if the x-ray pulse from LCLS is focused to a spot a few microns across. At this intensity there is no observable resonant emission, and photoionization essentially dominates at photon energies beyond 7400~eV.

Increasing the peak intensity by a factor of ten produces far more ionization and shows a significant difference in x-ray emission intensity between different CL models, of just under a factor two. This seems to be a systematic trend and is well within the capabilities of an experiment to resolve. The reason for the difference is that in the SP model both K-shell photoionization and $1s-3p$ photoexcitation can take place, while the EK model depresses the ionization potential to the point where $n=3$ states are not always present. The total emission intensity is thus a (crude) measure of the number of states able to participate in the absorption and emission processes. The total emission above 7.1~keV grows slower in SP simulations, mimicking the trend observed in Al, and reaches a higher peak at a higher photon energy. The peak is shifted compared with the EK simulations because the ionization thresholds are higher in the SP model, and is higher because the states are overall more numerous. The photon energy at which this inversion occurs is strongly correlated to the degree of CL, and is thus an information-rich quantity that is straightforward to determine experimentally.

We also calculate the emission under x-ray irradiation intensities of $10^{19}$~Wcm$^{-2}$, achievable by focusing the FEL beam to spot sizes of order 100~nm. Here the intensity -- and hence deposited energy density -- is sufficiently high to involve most if not all Fe charge states, increasing the total emission across the board, but most significantly at higher photon energies. The intensity is also sufficiently high to drive a significant number of double core hole transitions, placing most of the data points on the figure within our mixed region of competing photoionization and photoexcitation interactions. With a few exceptions (i.e., the resonances, and the region around 7600~eV), this high intensity regime seems less suitable for discriminating between CL models, but could be used in conjunction with lower intensity data to help further constrain models.

\section{Conclusions}
We have shown that the total integrated x-ray intensity emanating from an HED system driven by a focused, narrow-bandwidth x-ray pulse, can be used to discriminate between, and validate, different models of continuum lowering.
Applying the method to simulations and experimental data for Al systems present in the literature yields the expected results with good fidelity. We have extended our calculations to the case of x-ray heated Fe, and describe the conditions best suited for experimental measurements detailing the underlying physics.
Our approach is advantageous because it does not rely on the identification of specific spectroscopic features and can thus be used more generally, even where there is significant overlap between emission from many different charge states, as is the case in high-Z plasmas.

The method should also be less sensitive to the accuracy of detailed atomic modelling than previous spectroscopic approaches, provided the overall electronic structure is correctly accounted for, and is experimentally far simpler to field. In addition to measuring the total emitted intensity, the method could be coupled with a measurement of the total absorbed x-ray fraction as discussed in ref.~\cite{Rackstraw2015}, to provide a comprehensive, if integrated, total energy balance measurement of the interaction. This would act to further constrain the physical processes governing the change in ionization potentials and pressure ionization.

The method exhibits a sensitivity to the x-ray intensity, which must therefore be well characterized experimentally. However, the method is only weakly sensitive to small variations in the intensity distribution on target. This is in stark contrast with methods based on detailed spectroscopy, where obtaining accurate measurements of the intensity distribution is of paramount importance. We believe this weak dependence will be most advantageous in experiments employing the tightest focal spots, i.e., in nanofocusing campaigns, where detailed spatial variations of the x-ray pulse intensity can be more challenging to accurately determine.

S.M.V. gratefully acknowledges support from the Royal Society.
M.F.K., J.S.W and S.M.V. acknowledge support from the U.K. EPSRC under grant EP/P015794/1.


\begin{thebibliography}{31}%
\makeatletter
\providecommand \@ifxundefined [1]{%
 \@ifx{#1\undefined}
}%
\providecommand \@ifnum [1]{%
 \ifnum #1\expandafter \@firstoftwo
 \else \expandafter \@secondoftwo
 \fi
}%
\providecommand \@ifx [1]{%
 \ifx #1\expandafter \@firstoftwo
 \else \expandafter \@secondoftwo
 \fi
}%
\providecommand \natexlab [1]{#1}%
\providecommand \enquote  [1]{``#1''}%
\providecommand \bibnamefont  [1]{#1}%
\providecommand \bibfnamefont [1]{#1}%
\providecommand \citenamefont [1]{#1}%
\providecommand \href@noop [0]{\@secondoftwo}%
\providecommand \href [0]{\begingroup \@sanitize@url \@href}%
\providecommand \@href[1]{\@@startlink{#1}\@@href}%
\providecommand \@@href[1]{\endgroup#1\@@endlink}%
\providecommand \@sanitize@url [0]{\catcode `\\12\catcode `\$12\catcode
  `\&12\catcode `\#12\catcode `\^12\catcode `\_12\catcode `\%12\relax}%
\providecommand \@@startlink[1]{}%
\providecommand \@@endlink[0]{}%
\providecommand \url  [0]{\begingroup\@sanitize@url \@url }%
\providecommand \@url [1]{\endgroup\@href {#1}{\urlprefix }}%
\providecommand \urlprefix  [0]{URL }%
\providecommand \Eprint [0]{\href }%
\providecommand \doibase [0]{http://dx.doi.org/}%
\providecommand \selectlanguage [0]{\@gobble}%
\providecommand \bibinfo  [0]{\@secondoftwo}%
\providecommand \bibfield  [0]{\@secondoftwo}%
\providecommand \translation [1]{[#1]}%
\providecommand \BibitemOpen [0]{}%
\providecommand \bibitemStop [0]{}%
\providecommand \bibitemNoStop [0]{.\EOS\space}%
\providecommand \EOS [0]{\spacefactor3000\relax}%
\providecommand \BibitemShut  [1]{\csname bibitem#1\endcsname}%
\let\auto@bib@innerbib\@empty
%</preamble>
\bibitem [{\citenamefont {Starrett}\ and\ \citenamefont
  {Saumon}(2013)}]{PhysRevE.87.013104}%
  \BibitemOpen
  \bibfield  {author} {\bibinfo {author} {\bibfnamefont {C.~E.}\ \bibnamefont
  {Starrett}}\ and\ \bibinfo {author} {\bibfnamefont {D.}~\bibnamefont
  {Saumon}},\ }\href {\doibase 10.1103/PhysRevE.87.013104} {\bibfield
  {journal} {\bibinfo  {journal} {Phys. Rev. E}\ }\textbf {\bibinfo {volume}
  {87}},\ \bibinfo {pages} {013104} (\bibinfo {year} {2013})}\BibitemShut
  {NoStop}%
\bibitem [{\citenamefont {Das}\ \emph {et~al.}(2016)\citenamefont {Das},
  \citenamefont {Sahoo},\ and\ \citenamefont {Pal}}]{PhysRevA.93.052513}%
  \BibitemOpen
  \bibfield  {author} {\bibinfo {author} {\bibfnamefont {M.}~\bibnamefont
  {Das}}, \bibinfo {author} {\bibfnamefont {B.~K.}\ \bibnamefont {Sahoo}}, \
  and\ \bibinfo {author} {\bibfnamefont {S.}~\bibnamefont {Pal}},\ }\href
  {\doibase 10.1103/PhysRevA.93.052513} {\bibfield  {journal} {\bibinfo
  {journal} {Phys. Rev. A}\ }\textbf {\bibinfo {volume} {93}},\ \bibinfo
  {pages} {052513} (\bibinfo {year} {2016})}\BibitemShut {NoStop}%
\bibitem [{\citenamefont {Mondal}\ \emph {et~al.}(2013)\citenamefont {Mondal},
  \citenamefont {Dutta}, \citenamefont {Dixit},\ and\ \citenamefont
  {Majumder}}]{PhysRevA.87.062502}%
  \BibitemOpen
  \bibfield  {author} {\bibinfo {author} {\bibfnamefont {P.~K.}\ \bibnamefont
  {Mondal}}, \bibinfo {author} {\bibfnamefont {N.~N.}\ \bibnamefont {Dutta}},
  \bibinfo {author} {\bibfnamefont {G.}~\bibnamefont {Dixit}}, \ and\ \bibinfo
  {author} {\bibfnamefont {S.}~\bibnamefont {Majumder}},\ }\href {\doibase
  10.1103/PhysRevA.87.062502} {\bibfield  {journal} {\bibinfo  {journal} {Phys.
  Rev. A}\ }\textbf {\bibinfo {volume} {87}},\ \bibinfo {pages} {062502}
  (\bibinfo {year} {2013})}\BibitemShut {NoStop}%
\bibitem [{\citenamefont {Belkhiri}\ and\ \citenamefont
  {Poirier}(2013)}]{Belkhiri2013}%
  \BibitemOpen
  \bibfield  {author} {\bibinfo {author} {\bibfnamefont {M.}~\bibnamefont
  {Belkhiri}}\ and\ \bibinfo {author} {\bibfnamefont {M.}~\bibnamefont
  {Poirier}},\ }\href {\doibase 10.1016/j.hedp.2013.05.016} {\bibfield
  {journal} {\bibinfo  {journal} {HEDP}\ }\textbf {\bibinfo {volume} {9}},\
  \bibinfo {pages} {609} (\bibinfo {year} {2013})}\BibitemShut {NoStop}%
\bibitem [{\citenamefont {Crowley}(2014)}]{Crowley2014}%
  \BibitemOpen
  \bibfield  {author} {\bibinfo {author} {\bibfnamefont {B.}~\bibnamefont
  {Crowley}},\ }\href {\doibase http://dx.doi.org/10.1016/j.hedp.2014.04.003}
  {\bibfield  {journal} {\bibinfo  {journal} {HEDP}\ }\textbf {\bibinfo
  {volume} {13}},\ \bibinfo {pages} {84 } (\bibinfo {year} {2014})}\BibitemShut
  {NoStop}%
\bibitem [{\citenamefont {Calisti}\ \emph {et~al.}(2015)\citenamefont
  {Calisti}, \citenamefont {Ferri},\ and\ \citenamefont
  {Talin}}]{Calisti:2015}%
  \BibitemOpen
  \bibfield  {author} {\bibinfo {author} {\bibfnamefont {A.}~\bibnamefont
  {Calisti}}, \bibinfo {author} {\bibfnamefont {S.}~\bibnamefont {Ferri}}, \
  and\ \bibinfo {author} {\bibfnamefont {B.}~\bibnamefont {Talin}},\ }\href
  {http://stacks.iop.org/0953-4075/48/i=22/a=224003} {\bibfield  {journal}
  {\bibinfo  {journal} {J. Phys. B}\ }\textbf {\bibinfo {volume} {48}},\
  \bibinfo {pages} {224003} (\bibinfo {year} {2015})}\BibitemShut {NoStop}%
\bibitem [{\citenamefont {Stransky}(2016)}]{Stranski:2016}%
  \BibitemOpen
  \bibfield  {author} {\bibinfo {author} {\bibfnamefont {M.}~\bibnamefont
  {Stransky}},\ }\href {\doibase 10.1063/1.4940313} {\bibfield  {journal}
  {\bibinfo  {journal} {Phys. Plasmas}\ }\textbf {\bibinfo {volume} {23}},\
  \bibinfo {pages} {012708} (\bibinfo {year} {2016})}\BibitemShut {NoStop}%
\bibitem [{\citenamefont {Lin}\ \emph {et~al.}(2017)\citenamefont {Lin},
  \citenamefont {R\"opke}, \citenamefont {Kraeft},\ and\ \citenamefont
  {Reinholz}}]{PhysRevE.96.013202}%
  \BibitemOpen
  \bibfield  {author} {\bibinfo {author} {\bibfnamefont {C.}~\bibnamefont
  {Lin}}, \bibinfo {author} {\bibfnamefont {G.}~\bibnamefont {R\"opke}},
  \bibinfo {author} {\bibfnamefont {W.-D.}\ \bibnamefont {Kraeft}}, \ and\
  \bibinfo {author} {\bibfnamefont {H.}~\bibnamefont {Reinholz}},\ }\href
  {\doibase 10.1103/PhysRevE.96.013202} {\bibfield  {journal} {\bibinfo
  {journal} {Phys. Rev. E}\ }\textbf {\bibinfo {volume} {96}},\ \bibinfo
  {pages} {013202} (\bibinfo {year} {2017})}\BibitemShut {NoStop}%
\bibitem [{\citenamefont {Vinko}\ \emph {et~al.}(2014)\citenamefont {Vinko},
  \citenamefont {Ciricosta},\ and\ \citenamefont {Wark}}]{Vinko2014}%
  \BibitemOpen
  \bibfield  {author} {\bibinfo {author} {\bibfnamefont {S.~M.}\ \bibnamefont
  {Vinko}}, \bibinfo {author} {\bibfnamefont {O.}~\bibnamefont {Ciricosta}}, \
  and\ \bibinfo {author} {\bibfnamefont {J.~S.}\ \bibnamefont {Wark}},\
  }\href@noop {} {\bibfield  {journal} {\bibinfo  {journal} {Nat. Commun.}\
  }\textbf {\bibinfo {volume} {5}},\ \bibinfo {pages} {3533} (\bibinfo {year}
  {2014})}\BibitemShut {NoStop}%
\bibitem [{\citenamefont {Zhang}\ \emph {et~al.}(2016)\citenamefont {Zhang},
  \citenamefont {Zhao}, \citenamefont {Kang}, \citenamefont {Zhang},\ and\
  \citenamefont {He}}]{PhysRevB.93.115114}%
  \BibitemOpen
  \bibfield  {author} {\bibinfo {author} {\bibfnamefont {S.}~\bibnamefont
  {Zhang}}, \bibinfo {author} {\bibfnamefont {S.}~\bibnamefont {Zhao}},
  \bibinfo {author} {\bibfnamefont {W.}~\bibnamefont {Kang}}, \bibinfo {author}
  {\bibfnamefont {P.}~\bibnamefont {Zhang}}, \ and\ \bibinfo {author}
  {\bibfnamefont {X.-T.}\ \bibnamefont {He}},\ }\href {\doibase
  10.1103/PhysRevB.93.115114} {\bibfield  {journal} {\bibinfo  {journal} {Phys.
  Rev. B}\ }\textbf {\bibinfo {volume} {93}},\ \bibinfo {pages} {115114}
  (\bibinfo {year} {2016})}\BibitemShut {NoStop}%
\bibitem [{\citenamefont {Hu}(2017)}]{PhysRevLett.119.065001}%
  \BibitemOpen
  \bibfield  {author} {\bibinfo {author} {\bibfnamefont {S.~X.}\ \bibnamefont
  {Hu}},\ }\href {\doibase 10.1103/PhysRevLett.119.065001} {\bibfield
  {journal} {\bibinfo  {journal} {Phys. Rev. Lett.}\ }\textbf {\bibinfo
  {volume} {119}},\ \bibinfo {pages} {065001} (\bibinfo {year}
  {2017})}\BibitemShut {NoStop}%
\bibitem [{\citenamefont {Fletcher}\ \emph {et~al.}(2014)\citenamefont
  {Fletcher} \emph {et~al.}}]{Fletcher2014}%
  \BibitemOpen
  \bibfield  {author} {\bibinfo {author} {\bibfnamefont {L.~B.}\ \bibnamefont
  {Fletcher}} \emph {et~al.},\ }\href {\doibase 10.1103/PhysRevLett.112.145004}
  {\bibfield  {journal} {\bibinfo  {journal} {Phys. Rev. Lett.}\ }\textbf
  {\bibinfo {volume} {112}},\ \bibinfo {pages} {145004} (\bibinfo {year}
  {2014})}\BibitemShut {NoStop}%
\bibitem [{\citenamefont {Kraus}\ \emph {et~al.}(2016)\citenamefont {Kraus}
  \emph {et~al.}}]{PhysRevE.94.011202}%
  \BibitemOpen
  \bibfield  {author} {\bibinfo {author} {\bibfnamefont {D.}~\bibnamefont
  {Kraus}} \emph {et~al.},\ }\href {\doibase 10.1103/PhysRevE.94.011202}
  {\bibfield  {journal} {\bibinfo  {journal} {Phys. Rev. E}\ }\textbf {\bibinfo
  {volume} {94}},\ \bibinfo {pages} {011202} (\bibinfo {year}
  {2016})}\BibitemShut {NoStop}%
\bibitem [{\citenamefont {Ciricosta}\ \emph {et~al.}(2012)\citenamefont
  {Ciricosta} \emph {et~al.}}]{Ciricosta2012}%
  \BibitemOpen
  \bibfield  {author} {\bibinfo {author} {\bibfnamefont {O.}~\bibnamefont
  {Ciricosta}} \emph {et~al.},\ }\href {\doibase
  10.1103/PhysRevLett.109.065002} {\bibfield  {journal} {\bibinfo  {journal}
  {Phys. Rev. Lett.}\ }\textbf {\bibinfo {volume} {109}},\ \bibinfo {pages}
  {065002} (\bibinfo {year} {2012})}\BibitemShut {NoStop}%
\bibitem [{\citenamefont {Ciricosta}\ \emph
  {et~al.}(2016{\natexlab{a}})\citenamefont {Ciricosta} \emph
  {et~al.}}]{Ciricosta:2016aa}%
  \BibitemOpen
  \bibfield  {author} {\bibinfo {author} {\bibfnamefont {O.}~\bibnamefont
  {Ciricosta}} \emph {et~al.},\ }\href {http://dx.doi.org/10.1038/ncomms11713}
  {\bibfield  {journal} {\bibinfo  {journal} {Nat. Commun.}\ }\textbf {\bibinfo
  {volume} {7}},\ \bibinfo {pages} {11713} (\bibinfo {year}
  {2016}{\natexlab{a}})}\BibitemShut {NoStop}%
\bibitem [{\citenamefont {Preston}\ \emph {et~al.}(2017)\citenamefont {Preston}
  \emph {et~al.}}]{PhysRevLett.119.085001}%
  \BibitemOpen
  \bibfield  {author} {\bibinfo {author} {\bibfnamefont {T.~R.}\ \bibnamefont
  {Preston}} \emph {et~al.},\ }\href {\doibase 10.1103/PhysRevLett.119.085001}
  {\bibfield  {journal} {\bibinfo  {journal} {Phys. Rev. Lett.}\ }\textbf
  {\bibinfo {volume} {119}},\ \bibinfo {pages} {085001} (\bibinfo {year}
  {2017})}\BibitemShut {NoStop}%
\bibitem [{\citenamefont {Hoarty}\ \emph {et~al.}(2013)\citenamefont {Hoarty}
  \emph {et~al.}}]{Hoarty2013}%
  \BibitemOpen
  \bibfield  {author} {\bibinfo {author} {\bibfnamefont {D.~J.}\ \bibnamefont
  {Hoarty}} \emph {et~al.},\ }\href {\doibase 10.1103/PhysRevLett.110.265003}
  {\bibfield  {journal} {\bibinfo  {journal} {Phys. Rev. Lett.}\ }\textbf
  {\bibinfo {volume} {110}},\ \bibinfo {pages} {265003} (\bibinfo {year}
  {2013})}\BibitemShut {NoStop}%
\bibitem [{\citenamefont {Preston}\ \emph {et~al.}(2013)\citenamefont
  {Preston}, \citenamefont {Vinko}, \citenamefont {Ciricosta}, \citenamefont
  {Chung}, \citenamefont {Lee},\ and\ \citenamefont {Wark}}]{Preston2013}%
  \BibitemOpen
  \bibfield  {author} {\bibinfo {author} {\bibfnamefont {T.}~\bibnamefont
  {Preston}}, \bibinfo {author} {\bibfnamefont {S.}~\bibnamefont {Vinko}},
  \bibinfo {author} {\bibfnamefont {O.}~\bibnamefont {Ciricosta}}, \bibinfo
  {author} {\bibfnamefont {H.-K.}\ \bibnamefont {Chung}}, \bibinfo {author}
  {\bibfnamefont {R.}~\bibnamefont {Lee}}, \ and\ \bibinfo {author}
  {\bibfnamefont {J.}~\bibnamefont {Wark}},\ }\href@noop {} {\bibfield
  {journal} {\bibinfo  {journal} {HEDP}\ }\textbf {\bibinfo {volume} {9}},\
  \bibinfo {pages} {258} (\bibinfo {year} {2013})}\BibitemShut {NoStop}%
\bibitem [{\citenamefont {Stewart}\ and\ \citenamefont
  {Pyatt}(1966)}]{Stewart:1966ay}%
  \BibitemOpen
  \bibfield  {author} {\bibinfo {author} {\bibfnamefont {J.~C.}\ \bibnamefont
  {Stewart}}\ and\ \bibinfo {author} {\bibfnamefont {K.~J.}\ \bibnamefont
  {Pyatt}},\ }\href@noop {} {\bibfield  {journal} {\bibinfo  {journal}
  {Astrophys. J.}\ }\textbf {\bibinfo {volume} {144}},\ \bibinfo {pages} {1203}
  (\bibinfo {year} {1966})}\BibitemShut {NoStop}%
\bibitem [{\citenamefont {Vinko}\ \emph {et~al.}(2012)\citenamefont {Vinko}
  \emph {et~al.}}]{Vinko2012}%
  \BibitemOpen
  \bibfield  {author} {\bibinfo {author} {\bibfnamefont {S.~M.}\ \bibnamefont
  {Vinko}} \emph {et~al.},\ }\href {\doibase 10.1038/nature10746} {\bibfield
  {journal} {\bibinfo  {journal} {Nature}\ }\textbf {\bibinfo {volume} {482}},\
  \bibinfo {pages} {59} (\bibinfo {year} {2012})}\BibitemShut {NoStop}%
\bibitem [{\citenamefont {Cho}\ \emph {et~al.}(2012)\citenamefont {Cho} \emph
  {et~al.}}]{Cho2012}%
  \BibitemOpen
  \bibfield  {author} {\bibinfo {author} {\bibfnamefont {B.}~\bibnamefont
  {Cho}} \emph {et~al.},\ }\href {\doibase 10.1103/PhysRevLett.109.245003}
  {\bibfield  {journal} {\bibinfo  {journal} {Phys. Rev. Lett.}\ }\textbf
  {\bibinfo {volume} {109}},\ \bibinfo {pages} {245003} (\bibinfo {year}
  {2012})}\BibitemShut {NoStop}%
\bibitem [{\citenamefont {Ciricosta}\ \emph
  {et~al.}(2016{\natexlab{b}})\citenamefont {Ciricosta}, \citenamefont {Vinko},
  \citenamefont {Chung}, \citenamefont {Jackson}, \citenamefont {Lee},
  \citenamefont {Preston}, \citenamefont {Rackstraw},\ and\ \citenamefont
  {Wark}}]{Ciricosta:2016POP}%
  \BibitemOpen
  \bibfield  {author} {\bibinfo {author} {\bibfnamefont {O.}~\bibnamefont
  {Ciricosta}}, \bibinfo {author} {\bibfnamefont {S.~M.}\ \bibnamefont
  {Vinko}}, \bibinfo {author} {\bibfnamefont {H.-K.}\ \bibnamefont {Chung}},
  \bibinfo {author} {\bibfnamefont {C.}~\bibnamefont {Jackson}}, \bibinfo
  {author} {\bibfnamefont {R.~W.}\ \bibnamefont {Lee}}, \bibinfo {author}
  {\bibfnamefont {T.~R.}\ \bibnamefont {Preston}}, \bibinfo {author}
  {\bibfnamefont {D.~S.}\ \bibnamefont {Rackstraw}}, \ and\ \bibinfo {author}
  {\bibfnamefont {J.~S.}\ \bibnamefont {Wark}},\ }\href
  {http://scitation.aip.org/content/aip/journal/pop/23/2/10.1063/1.4942540}
  {\bibfield  {journal} {\bibinfo  {journal} {Phys. Plasmas}\ }\textbf
  {\bibinfo {volume} {23}},\ \bibinfo {eid} {022707} (\bibinfo {year}
  {2016}{\natexlab{b}})}\BibitemShut {NoStop}%
\bibitem [{\citenamefont {Vinko}(2015)}]{Vinko:jpp2015}%
  \BibitemOpen
  \bibfield  {author} {\bibinfo {author} {\bibfnamefont {S.~M.}\ \bibnamefont
  {Vinko}},\ }\href@noop {} {\bibfield  {journal} {\bibinfo  {journal} {Journal
  of Plasma Physics}\ }\textbf {\bibinfo {volume} {81}},\ \bibinfo {pages}
  {365810501} (\bibinfo {year} {2015})}\BibitemShut {NoStop}%
\bibitem [{\citenamefont {Vinko}\ \emph {et~al.}(2015)\citenamefont {Vinko}
  \emph {et~al.}}]{Vinko2015}%
  \BibitemOpen
  \bibfield  {author} {\bibinfo {author} {\bibfnamefont {S.~M.}\ \bibnamefont
  {Vinko}} \emph {et~al.},\ }\href {http://dx.doi.org/10.1038/ncomms7397}
  {\bibfield  {journal} {\bibinfo  {journal} {Nat Commun}\ }\textbf {\bibinfo
  {volume} {6}},\ \bibinfo {pages} {6397} (\bibinfo {year} {2015})}\BibitemShut
  {NoStop}%
\bibitem [{\citenamefont {van~den Berg}\ \emph {et~al.}(2017)\citenamefont
  {van~den Berg} \emph {et~al.}}]{quincy:2017}%
  \BibitemOpen
  \bibfield  {author} {\bibinfo {author} {\bibfnamefont {Q.}~\bibnamefont
  {van~den Berg}} \emph {et~al.},\ }\href@noop {} {\bibfield  {journal}
  {\bibinfo  {journal} {Accepted for publication in Phys. Rev. Lett.}\ }
  (\bibinfo {year} {2017})}\BibitemShut {NoStop}%
\bibitem [{\citenamefont {Chung}\ \emph {et~al.}(2007)\citenamefont {Chung},
  \citenamefont {Chen},\ and\ \citenamefont {Lee}}]{SCFLYcode}%
  \BibitemOpen
  \bibfield  {author} {\bibinfo {author} {\bibfnamefont {H.-K.}\ \bibnamefont
  {Chung}}, \bibinfo {author} {\bibfnamefont {M.~H.}\ \bibnamefont {Chen}}, \
  and\ \bibinfo {author} {\bibfnamefont {R.~W.}\ \bibnamefont {Lee}},\
  }\href@noop {} {\bibfield  {journal} {\bibinfo  {journal} {High Energy
  Density Physics}\ }\textbf {\bibinfo {volume} {3}},\ \bibinfo {pages} {57}
  (\bibinfo {year} {2007})}\BibitemShut {NoStop}%
\bibitem [{\citenamefont {Ecker}\ and\ \citenamefont
  {Kr{\"o}ll}(1963)}]{EckerG.Kroll1963}%
  \BibitemOpen
  \bibfield  {author} {\bibinfo {author} {\bibnamefont {Ecker}}\ and\ \bibinfo
  {author} {\bibnamefont {Kr{\"o}ll}},\ }\href {\doibase 10.1063/1.1724509}
  {\bibfield  {journal} {\bibinfo  {journal} {Phys. Fluids}\ }\textbf {\bibinfo
  {volume} {6}},\ \bibinfo {pages} {62} (\bibinfo {year} {1963})}\BibitemShut
  {NoStop}%
\bibitem [{\citenamefont {Iglesias}(2014)}]{Iglesias2014}%
  \BibitemOpen
  \bibfield  {author} {\bibinfo {author} {\bibfnamefont {C.~A.}\ \bibnamefont
  {Iglesias}},\ }\href {\doibase http://dx.doi.org/10.1016/j.hedp.2014.04.002}
  {\bibfield  {journal} {\bibinfo  {journal} {HEDP}\ }\textbf {\bibinfo
  {volume} {12}},\ \bibinfo {pages} {5 } (\bibinfo {year} {2014})}\BibitemShut
  {NoStop}%
\bibitem [{\citenamefont {Chalupsk\'{y}}\ \emph {et~al.}(2010)\citenamefont
  {Chalupsk\'{y}} \emph {et~al.}}]{Chalupsky:2010bv}%
  \BibitemOpen
  \bibfield  {author} {\bibinfo {author} {\bibfnamefont {J.}~\bibnamefont
  {Chalupsk\'{y}}} \emph {et~al.},\ }\href
  {http://www.ncbi.nlm.nih.gov/pubmed/21197057} {\bibfield  {journal} {\bibinfo
   {journal} {Opt. Express}\ }\textbf {\bibinfo {volume} {18}},\ \bibinfo
  {pages} {27836} (\bibinfo {year} {2010})}\BibitemShut {NoStop}%
\bibitem [{\citenamefont {Chalupsk\'{y}}\ \emph {et~al.}(2013)\citenamefont
  {Chalupsk\'{y}} \emph {et~al.}}]{Chalupsky:2013}%
  \BibitemOpen
  \bibfield  {author} {\bibinfo {author} {\bibfnamefont {J.}~\bibnamefont
  {Chalupsk\'{y}}} \emph {et~al.},\ }\href@noop {} {\bibfield  {journal}
  {\bibinfo  {journal} {Opt. Express}\ }\textbf {\bibinfo {volume} {21}},\
  \bibinfo {pages} {26363} (\bibinfo {year} {2013})}\BibitemShut {NoStop}%
\bibitem [{\citenamefont {Rackstraw}\ \emph {et~al.}(2015)\citenamefont
  {Rackstraw} \emph {et~al.}}]{Rackstraw2015}%
  \BibitemOpen
  \bibfield  {author} {\bibinfo {author} {\bibfnamefont {D.}~\bibnamefont
  {Rackstraw}} \emph {et~al.},\ }\href {\doibase
  10.1103/PhysRevLett.114.015003} {\bibfield  {journal} {\bibinfo  {journal}
  {Phys. Rev. Lett.}\ }\textbf {\bibinfo {volume} {114}},\ \bibinfo {pages}
  {015003} (\bibinfo {year} {2015})}\BibitemShut {NoStop}%
\end{thebibliography}
\end{document}